\documentstyle[12pt,epsf,axodraw]{article} \textwidth 16.5cm
\newcommand{\beq}[1]{
\begin{equation}\label{#1}}
\newcommand{\eeq}{\end{equation}}
\newcommand{\bea}[1]{
\begin{eqnarray}\label{#1}}
\newcommand{\eea}{\end{eqnarray}}

\begin{document}

\begin{titlepage}

\begin{flushright}
CERN--PH--TH/2007--128
\end{flushright}

\begin{center}
{\LARGE \bf Collinear improvement of the BFKL kernel in the electroproduction 
of two light vector mesons}

\vspace{1cm}

{\sc F. Caporale}${}^{1}$, {\sc A.~Papa}${}^{1}$ and {\sc A.~Sabio~Vera}${}^{2}$
\\[0.5cm]
\vspace*{0.1cm}
${}^1$ {\it Dipartimento di Fisica, Universit\`a
della Calabria \\
and Istituto Nazionale di Fisica Nucleare, Gruppo collegato di Cosenza \\
I-87036 Arcavacata di Rende, Cosenza, Italy} \\[0.2cm]
${}^2$ {\it Physics Department, Theory Division, CERN,}\\
{\it CH--1211, Geneva 23, Switzerland}\\[1.0cm]
\vspace*{0.5cm}

\centerline{\large \em \today}

\vskip2cm
{\bf Abstract\\[10pt]} \parbox[t]{\textwidth}
{The use of the BFKL kernel improved by the inclusion of subleading terms generated by 
renormalization group (RG) analysis has been suggested to cure the instabilities in
the behavior of the BFKL Green's function in the next-to-leading approximation (NLA).
We test the performance of a RG-improved kernel in the determination of the
amplitude of a physical process, the electroproduction of two light vector mesons,
in the BFKL approach in the NLA. We find that a smooth behavior of the amplitude 
with the center-of-mass energy can be achieved, setting the renormalization and 
energy scales appearing in the subleading terms to values much closer to the 
kinematical scales of the process than in the approaches based on the 
unimproved kernel.}

\vskip1cm
\end{center}

\vspace*{1cm}

\end{titlepage}

\section{Introduction}

\begin{figure}[tb]
\centering
\setlength{\unitlength}{0.35mm}
\begin{picture}(300,200)(0,0)

\Photon(0,190)(100,190){3}{7}
\ArrowLine(200,190)(300,190)
\Text(50,200)[c]{$p$}
\Text(250,200)[c]{$p_1$}
\Text(150,190)[]{$\Phi_1(\vec q_1, s_0)$}
\Oval(150,190)(20,50)(0)

\ZigZag(125,174)(125,120){5}{7}
\ZigZag(175,174)(175,120){5}{7}
\ZigZag(125,26)(125,80){5}{7}
\ZigZag(175,26)(175,80){5}{7}

\ArrowLine(110,160)(110,130)
\ArrowLine(190,130)(190,160)
\ArrowLine(110,70)(110,40)
\ArrowLine(190,40)(190,70)

\Text(100,145)[r]{$q_1$}
\Text(200,145)[l]{$q_1$}
\Text(100,55)[r]{$q_2$}
\Text(200,55)[l]{$q_2$}

\GCirc(150,100){30}{1}
\Text(150,100)[]{$G(\vec q_1,\vec q_2)$}

\Photon(0,10)(100,10){3}{7}
\ArrowLine(200,10)(300,10)
\Text(50,0)[c]{$p'$}
\Text(250,0)[c]{$p_2$}
\Text(150,10)[]{$\Phi_2(-\vec q_2,s_0)$}
\Oval(150,10)(20,50)(0)

\end{picture}

\caption[]{Schematic representation of the amplitude for the $\gamma^*(p)\,
\gamma^*(p') \to V(p_1)\, V(p_2)$ forward scattering.}
\label{fig:BFKL}
\end{figure}
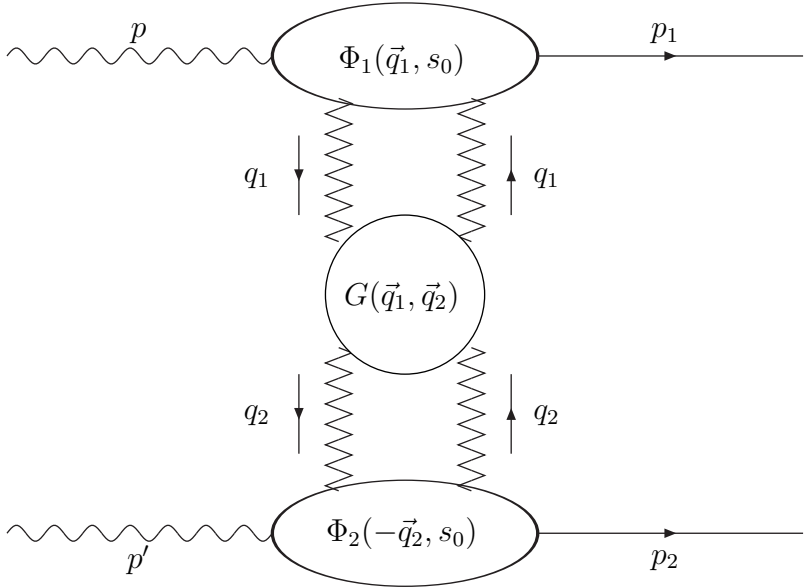

It is known that hard processes in which the center-of-mass energy is much 
larger than all the other scales are the natural ground for the application of
the BFKL approach~\cite{BFKL}. This approach was originally developed in the 
leading logarithmic approximation (LLA), which means resummation of all terms of 
the form $(\alpha_s \ln(s))^n$. In such an approximation the argument $\mu_R$ 
of the running coupling and the energy scale are not fixed. This motivated 
the extension of the approach to the next-to-leading logarithmic approximation 
(NLLA), which means resummation of all terms proportional to 
$\alpha_s(\alpha_s \ln(s))^n$. In both approximations the BFKL amplitude  
appears as a convolution of the Green's function of two interacting Reggeized 
gluons with the impact factors of the colliding particles (see, for example,
Fig.~\ref{fig:BFKL}).
The Green's function, which carries the dependence on the center-of-mass energy, 
can be determined through the BFKL equation. The impact factors are 
process-dependent and describe the interaction between Reggeized gluons and 
scattering particles.

The singlet kernel of the BFKL equation in the next-to-leading approximation (NLA) was obtained for the forward case in Ref.~\cite{NLA-kernel}, completing the long program of 
calculation of the NLA corrections~\cite{NLA-corrections} (for a review, see 
Ref.~\cite{news}).
In the non-forward case the ingredients for the NLA BFKL kernel have been 
known for a few years in the case of the color octet representation in the
$t$-channel~\cite{NLA-corrections-nf}. This color representation is very
important to check the consistency of the $s$-channel unitarity with
the gluon Reggeization, i.e. for the ``bootstrap''~\cite{bootstrap}.
More recently, the last missing piece for the determination
of the non-forward NLA BFKL kernel has been calculated 
in the singlet color representation, i.e.
in the Pomeron channel, relevant for physical applications~\cite{FF05}.
The singlet NLA BFKL kernel in the so-called ``dipole form'' is available
now also in the coordinate representation~\cite{coord}, which allows the study of
its conformal properties and the comparison with the kernel of the 
Balitsky-Kovchegov~\cite{BK} equation in the linear regime. So far, the color dipole
kernel has been calculated in the NLA only for the quark 
part~\cite{Balitsky:2006wa} and agrees with the dipole form of the quark part of 
the NLA BFKL kernel.

In this paper we will focus on the BFKL approach in the NLA and in 
the case of forward scattering. It is well known that the NLA corrections to the 
Green's function turn out to be large, this being a signal of the poor convergence of the BFKL series. In order to ``cure" the resulting instability, more convergent 
kernels have been introduced, including terms generated by renormalization 
group (RG), or collinear, analysis~\cite{Salam}. They are based on the 
$\omega$-shift method~\cite{Salam}, with $\omega$ being the variable Mellin-conjugated 
to the squared center-of-mass energy $s$. The main effect of this method is that 
the scale-invariant part of the kernel eigenvalues carries a dependence on the Mellin 
variable $\omega$, in such a way that the position of the singularities of the Green's function 
in the $\omega$-plane becomes the solution of an implicit equation in $\omega$. 
Many other studies have been performed, either based on this kind of improved 
kernels~\cite{Ciafalonietal} or analyzing different aspects of the kernel NLA and 
alternative approaches~\cite{NLOpapers}. The effects of these collinear 
corrections in exclusive observables have been investigated in Ref.~\cite{Jets},
with a posteriori confirmation in Ref.~\cite{Royon}.

In Ref.~\cite{SabioBess} the original approach of Ref.~\cite{Salam} was   
revisited and an approximation to the original $\omega$-shift was performed,
leading to an explicit expression for the RG-improved NLA kernel. It was  
shown that this improved kernel leads to a NLA BFKL 
Green's function exempt of instabilities. Since the effect of the 
RG-improvement is to modify the BFKL kernel by the inclusion of terms beyond the
NLA, one is led to conclude that RG-generated terms, although formally 
subleading, play an important numerical role in practical applications.

It is very interesting to test the RG-improvement of the kernel 
in the calculation of a full physical amplitude, rather than just considering
its effect on the BFKL Green's function, and to compare it with other approaches.
A test-field for this comparison can be provided by the physical process  
$\gamma^* \gamma^* \to VV$, where $\gamma^*$ represents a virtual photon and $V$ a 
light neutral vector meson ($\rho^0, \omega, \phi$).
The amplitude of this reaction has been calculated in Ref.~\cite{AlexDima1} through 
the convolution of the (unimproved) BFKL Green's function with the $\gamma^* \to 
V$ impact factors, calculated in Ref.~\cite{IKP04} \footnote{This amplitude
has been considered also in~\cite{Segond:2007fj,Enberg:2005eq,Pire:2005ic}.}. 
In the case of equal photon virtualities, the so-called ``pure" BFKL regime, a 
numerical calculation has shown that NLA corrections are large and of opposite 
sign with respect to the leading order and are dominated, at the lower energies, 
by the NLA corrections from the impact factors. Nonetheless, an amplitude for this
process with a smooth behavior in $s$ could be achieved by ``optimizing'' the
choice of the energy scale $s_0$ and of the renormalization scale $\mu_R$, which 
appear in the subleading terms. Later on it has been found that the result 
is rather stable under change of the method of optimization of the perturbative 
series and of the representation adopted for the amplitude~\cite{AlexDima2}. 

The striking feature of these investigations was that in all cases the optimal 
values of the two energy parameters turned out to be quite far from 
the kinematical scales of the reaction. For example, the optimal value of the
renormalization scale $\mu_R$ turned out to be typically as large as $\sim 10Q$,
$Q^2$ being the virtuality of the colliding photons. The proposed explanation for 
these ``unnatural" values was that they mimic the unknown next-to-NLA corrections, 
which should be large and of opposite sign respect to the NLA in order to preserve 
the renorm- and energy scale invariance of the exact amplitude. If this explanation is correct and if the RG-improvement of the kernel catches 
the essential dynamics from subleading orders, then, by repeating the numerical 
determination of the $\gamma^* \gamma^* \to VV$ amplitude with the use of an 
RG-improved kernel, one should get more ``natural" values for the optimal choices
of the energy scales and, of course, results consistent with the previous 
determinations. In this work we address this question by calculating the NLA amplitude
of the $\gamma^* \gamma^* \to VV$ process in the BFKL approach with the RG-improved 
kernel of Ref.~\cite{SabioBess}, which can be straightforwardly implemented in the 
numerical set up of Refs.~\cite{AlexDima1,AlexDima2}.

The paper is organized as follows: in the next Section we repeat the
steps of Refs.~\cite{AlexDima1,AlexDima2} to build up the NLA amplitude in two 
representations, series and ``exponentiated'', which implement the RG-improved kernel 
of Ref.~\cite{SabioBess}; in Section~3 we numerically evaluate the amplitude, considering both the cases of colliding photons with the same
virtualities and with strongly ordered virtualities. We stress that in 
Refs.~\cite{AlexDima1,AlexDima2} only the case of equal photons' virtualities was 
considered; attempts to determine the amplitude for strongly
ordered virtualities were unsuccessful, due to the large instabilities met in the 
numerical analysis~\cite{AlexDima3}. We expect that the RG-improvement should be even more
effective in the latter case, since it was conceived to work in a kinematics with
strong asymmetry in the transverse momentum plane~\cite{Salam}.

\section{The NLA amplitude with the RG-improved Green's 
function~\protect\footnote{This Section follows closely Section~2 of the  
Refs.~\cite{AlexDima1,AlexDima2}, the only difference being the use of a  
modified BFKL kernel. The reader already familiar with the notation and the 
previous papers may prefer to go straight to the main formulas:
Eq.~(\ref{amplnlaE}) for the ``exponentiated'' representation,
Eq.~(\ref{series}) for the ``series'' representation of the amplitude
and Eq.~(\ref{bessel}) for the extra-term in the BFKL kernel eigenvalue.}}

We consider the production of two light vector mesons ($V=\rho^0, \omega, \phi$) in
the collision of two virtual photons,
\beq{process}
\gamma^*(p) \: \gamma^*(p')\to V(p_1) \:V(p_2) \;.
\eeq
Here, $p_1$ and $p_2$ are taken as Sudakov vectors satisfying $p_1^2=p_2^2=0$ and
$2(p_1 p_2)=s$; the virtual photon momenta are instead
\beq{kinphoton}
p =\alpha p_1-\frac{Q_1^2}{\alpha s} p_2 \;, \hspace{2cm}
p'=\alpha^\prime p_2-\frac{Q_2^2}{\alpha^\prime s} p_1 \;,
\eeq
so that the photon virtualities turn to be $p^2=-Q_1^2$ and $(p')^2=-Q_2^2$.
We consider the kinematics when
\beq{kin}
s\gg Q^2_{1,2}\gg \Lambda^2_{QCD} \, ,
\eeq
and
\beq{alphas}
\alpha=1+\frac{Q_2^2}{s}+{\cal O}(s^{-2})\, , \quad
\alpha^\prime =1+\frac{Q_1^2}{s}+{\cal O}(s^{-2})\, .
\eeq 
In this case the vector mesons are produced by longitudinally polarized photons in
the longitudinally polarized state~\cite{IKP04}. Other helicity amplitudes are
power suppressed, with a suppression factor $\sim m_V/Q_{1,2}$.
We will discuss here the amplitude of the forward scattering, i.e.
when the transverse momenta of the produced $V$ mesons are zero or 
when the variable $t=(p_1-p)^2$ takes its maximal value $t_0=-Q_1^2Q_2^2/s+{\cal
O}(s^{-2})$.

The forward amplitude in the BFKL approach may be presented as follows
\beq{imA}
{\cal I}m_s\left( {\cal A} \right)=\frac{s}{(2\pi)^2}\int\frac{d^2\vec q_1}{\vec
q_1^{\,\, 2}}\Phi_1(\vec q_1,s_0)\int
\frac{d^2\vec q_2}{\vec q_2^{\,\,2}} \Phi_2(-\vec q_2,s_0)
\int\limits^{\delta +i\infty}_{\delta
-i\infty}\frac{d\omega}{2\pi i}\left(\frac{s}{s_0}\right)^\omega
G_\omega (\vec q_1, \vec q_2)\, .
\eeq
This representation for the amplitude is valid with NLA accuracy.

In Eq.~(\ref{imA}), $\Phi_{1}(\vec q_1,s_0)$ and $\Phi_{2}(-\vec q_2,s_0)$
are the impact factors describing the transitions $\gamma^*(p)\to V(p_1)$
and $\gamma^*(p')\to V(p_2)$, respectively. 
The Green's function in (\ref{imA}) is determined by the BFKL equation
\beq{Green}
\delta^2(\vec q_1-\vec q_2)=\omega \, G_\omega (\vec q_1, \vec q_2)-
\int d^2 \vec q \, K(\vec q_1,\vec q)\, G_\omega (\vec q, \vec q_2) \;,
\eeq
where $K(\vec q_1,\vec q_2)$ is the BFKL kernel.
It is convenient to work in the transverse momentum representation,
where ``transverse'' refers to the plane orthogonal to the vector mesons  
momenta. In this representation, defined by
\beq{transv}
\hat{\vec q}\: |\vec q_i\rangle = \vec q_i|\vec q_i\rangle\;,
\eeq
\beq{norm}
\langle\vec q_1|\vec q_2\rangle =\delta^{(2)}(\vec q_1 - \vec q_2) \;,
\hspace{2cm}
\langle A|B\rangle =
\langle A|\vec k\rangle\langle\vec k|B\rangle =
\int d^2k A(\vec k)B(\vec k)\;,
\eeq
the kernel of the operator $\hat K$ is
\beq{kernel-op}
K(\vec q_2, \vec q_1) = \langle\vec q_2| \hat K |\vec q_1\rangle
\eeq
and the equation for the Green's function reads 
\beq{Groper}
\hat 1=(\omega-\hat K)\hat G_\omega\;,
\eeq
its solution being
\beq{Groper1}
\hat G_\omega=(\omega-\hat K)^{-1} \, .
\eeq
To clearly indicate the RG-improved pieces of the kernel, 
we decompose $\hat K$ as
\beq{kern}
\hat K=\bar \alpha_s \hat K^0 + \bar \alpha_s^2 \hat K^1 +  \hat K_{RG} \;,
\eeq
where
\beq{baral}
{\bar \alpha_s}=\frac{\alpha_s N_c}{\pi}
\eeq
and $N_c$ is the number of colors. In Eq.~(\ref{kern}) $\hat K^0$ is the
BFKL kernel in the LLA, $\hat K^1$ is the NLA correction and $\hat K_{RG}$ 
includes the RG-generated terms, which are ${\cal O}({\bar \alpha_s^3})$.
The impact factors are also presented as an expansion in $\alpha_s$
\beq{impE}
\Phi_{1,2}(\vec q)= \alpha_s \,
D_{1,2}\left[C^{(0)}_{1,2}(\vec q^{\,\, 2})+\bar\alpha_s
C^{(1)}_{1,2}(\vec
q^{\,\, 2})\right] \, , \quad D_{1,2}=-\frac{4\pi e_q  f_V}{N_c Q_{1,2}}
\sqrt{N_c^2-1}\, ,
\eeq
where $f_V$ is the meson dimensional coupling constant ($f_{\rho}\approx
200\, \rm{ MeV}$) and $e_q$ should be replaced by $e/\sqrt{2}$, $e/(3\sqrt{2})$
and $-e/3$ for the case of $\rho^0$, $\omega$ and $\phi$ meson production,
respectively.

In the collinear factorization approach the meson transition impact factor
is given as a convolution of the hard scattering amplitude for the
production of a collinear quark--antiquark pair with the meson distribution
amplitude (DA). The integration variable in this convolution is the fraction $z$
of the meson momentum carried by the quark ($\bar z\equiv 1-z$ is
the momentum fraction carried by the antiquark):
\beq{imps1}
C^{(0)}_{1,2}(\vec q^{\,\, 2})=\int\limits^1_0 dz \,
\frac{\vec q^{\,\, 2}}{\vec q^{\,\, 2}+z \bar zQ_{1,2}^2}\phi_\parallel (z)
\, .
\eeq
The NLA correction to the hard scattering amplitude, for a photon with virtuality
equal to $Q^2$, is defined as follows
\beq{imps2}
C^{(1)}(\vec q^{\,\, 2})=\frac{1}{4 N_c}\int\limits^1_0 dz \,
\frac{\vec q^{\,\, 2}}{\vec q^{\,\, 2}+z \bar zQ^2}[\tau(z)+\tau(1-z)]
\phi_\parallel (z)
\, ,
\eeq
with $\tau(z)$ given in the Eq.~(75) of Ref.~\cite{IKP04}.
$C^{(1)}_{1,2}(\vec q^{\,\, 2})$ are given by the previous expression with
$Q^2$ replaced everywhere in the integrand by $Q^2_1$ and $Q^2_2$,
respectively. We will use the DA in the asymptotic form 
$\phi^{as}_\parallel(z)=6z(1-z)$.

To determine the amplitude with NLA accuracy we need an approximate
solution of Eq.~(\ref{Groper1}). With the required accuracy this solution
is
\beq{exp}
\hat G_\omega = (\omega-\bar \alpha_s\hat K^0)^{-1}+
(\omega-\bar \alpha_s\hat K^0)^{-1}\left(\bar \alpha_s^2 \hat K^1 
+ \hat K_{RG} \right) (\omega-\bar \alpha_s \hat K^0)^{-1}
+ {\cal O}\left[\left(\bar \alpha_s^2 \hat K^1\right)^2\right]
\, .
\eeq
Differently from Refs.~\cite{AlexDima1,AlexDima2}, where $\hat K_{RG}$ was absent,
this Green's function includes effects which are beyond the NLA.
The basis of eigenfunctions of the LLA kernel,
\beq{KLLA}
\hat K^0 |\nu\rangle = \chi(\nu)|\nu\rangle \, , \;\;\;\;\;\;\;\;\;\;
\chi (\nu)=
2\psi(1)-\psi\left(\frac{1}{2}+i\nu\right)-\psi\left(\frac{1}{2}-i\nu\right)\, ,
\eeq
is given by the following set of functions:
\beq{nuLLA}
\langle\vec q\, |\nu\rangle =\frac{1}{\pi \sqrt{2}}\left(\vec q^{\,\, 2}\right)
^{i\nu-\frac{1}{2}} \;,
\eeq
for which the  orthonormality  condition takes the form
\beq{ort}
\langle \nu^\prime | \nu\rangle =\int \frac{d^2\vec q}
{2 \pi^2 }\left(\vec q^{\,\, 2}\right)
^{i\nu-i\nu^\prime -1}=\delta(\nu-\nu^\prime)\, .
\eeq
The action of the modified BFKL kernel on these functions may be expressed
as follows:
\bea{Konnu}
\hat K|\nu\rangle &=&
\bar \alpha_s(\mu_R) \chi(\nu)|\nu\rangle
 +\bar \alpha_s^2(\mu_R)
\left(\chi^{(1)}(\nu)
+\frac{\beta_0}{4N_c}\chi(\nu)\ln(\mu^2_R)\right)|\nu\rangle
\nonumber \\
&+& \bar
\alpha_s^2(\mu_R)\frac{\beta_0}{4N_c}\chi(\nu)\left(i\frac{\partial}{\partial \nu}
\right)|\nu\rangle  + \chi_{RG} (\nu) |\nu\rangle \;,
\eea
where the first term represents the action of LLA kernel, the second
and the third ones stand for the diagonal and the non-diagonal parts of the
NLA BFKL kernel~\cite{AlexDima1} and 
\bea{bessel}
\chi_{RG} (\nu) &=& 2 \Re e \left\{\sum_{m=0}^{\infty} 
\left[\left(\sum_{n=0}^{\infty}\frac{(-1)^n (2n)!}{2^n n! (n+1)!}
\frac{\left({\bar \alpha}_s+ {\rm a} \,{\bar \alpha}_s^2\right)^{n+1}}
{\left(1/2 + i\nu + m - {\rm b} \,{\bar \alpha}_s\right)^{2n+1}}\right) 
\right. \right. \\
&-&  \left.\left. \frac{\bar{\alpha}_s}{1/2 + i\nu + m} - \bar{\alpha}_s^2 
\left(\frac{\rm a}{1/2 + i\nu +m} + \frac{\rm b}{(1/2 + i \nu + m)^2}
-\frac{1}{2(1/2 + i\nu+m)^3}\right)\right]\right\} \nonumber
\eea
is the solution of the $\omega$-shift equation obtained in \cite{SabioBess}, with 
\begin{eqnarray}
\label{ab}
{\rm a} &=& \frac{5}{12}\frac{\beta_0}{N_c} -\frac{13}{36}\frac{n_f}{N_c^3}
-\frac{55}{36}, \;\;\;\;\;
{\rm b} ~=~ -\frac{1}{8}\frac{\beta_0}{N_c} -\frac{n_f}{6 N_c^3}
-\frac{11}{12}.
\end{eqnarray}
The function $\chi^{(1)}(\nu)$ is conveniently represented in the form
\beq{ch11}
\chi^{(1)}(\nu)=-\frac{\beta_0}{8\, N_c}\left(
\chi^2(\nu)-\frac{10}{3}\chi(\nu)-i\chi^\prime(\nu)
\right) + {\bar \chi}(\nu)\, ,
\eeq
where
\bea{chibar}
\bar \chi(\nu)\,&=&\,-\frac{1}{4}\left[\frac{\pi^2-4}{3}\chi(\nu)-6\zeta(3)-
\chi^{\prime\prime}(\nu)-\frac{\pi^3}{\cosh(\pi\nu)}
\right.
\nonumber \\
&+& \left.
\frac{\pi^2\sinh(\pi\nu)}{2\,\nu\, \cosh^2(\pi\nu)}
\left(
3+\left(1+\frac{n_f}{N_c^3}\right)\frac{11+12\nu^2}{16(1+\nu^2)}
\right)
+\,4\,\phi(\nu)
\right] \, ,
\eea
\beq{phi}
\phi(\nu)\,=\,2\int\limits_0^1dx\,\frac{\cos(\nu\ln(x))}{(1+x)\sqrt{x}}
\left[\frac{\pi^2}{6}-\mbox{Li}_2(x)\right]\, , \;\;\;\;\;
\mbox{Li}_2(x)=-\int\limits_0^xdt\,\frac{\ln(1-t)}{t} \, .
\eeq
Here and below $\chi^\prime(\nu)=d( \chi (\nu) )/d\nu$ and $\chi^{\prime\prime}
(\nu)=d^2(\chi (\nu) )/d^2\nu$.

The $|\nu\rangle$ representations for the impact factors are given by the following 
expressions:
\beq{nuu}
\frac{C_1^{(0)}(\vec q^{\,\, 2})}{\vec
q^{\,\, 2}}=\int\limits_{-\infty}^{+\infty}\, d\, \nu^\prime \,c_1(\nu^\prime)
\langle\nu^\prime| \vec q\rangle \;, \hspace{2cm}
\frac{C_2^{(0)}(\vec q^{\,\, 2})}{\vec q^{\,\, 2}}=\int\limits_{-\infty}^{+\infty}
\, d\, \nu \,c_2(\nu)\,\langle\vec q|\nu\rangle \;,
\eeq
\beq{imp1}
c_1(\nu)=\int d^2\vec q \,\, C_1^{(0)}(\vec q^{\, 2})
\frac{\left(\vec q^{\, 2}\right)^{i\nu-\frac{3}{2}}}{\pi \sqrt{2}}
\, ,\;\;\;\;\;
c_2(\nu)=\int d^2\vec q \,\, C_2^{(0)}(\vec q^{\, 2})
\frac{\left(\vec q^{\, 2}\right)^{-i\nu-\frac{3}{2}}}{\pi \sqrt{2}} \, ,
\eeq
and by similar equations for $c_1^{(1)}(\nu)$ and $c_2^{(1)}(\nu)$
from the NLA corrections to the impact factors, $C_1^{(1)}(\vec
q^{\,\, 2})$ and $C_2^{(1)}(\vec q^{\,\, 2})$.

Following Ref.~\cite{AlexDima1}, we obtain the amplitude as a spectral 
decomposition on the basis of eigenfunctions of the LLA BFKL kernel:
\begin{eqnarray}
\frac{{\cal I}m_s\left( {\cal A} \right)}{D_1D_2} 
&=&\frac{s}{(2\pi)^2}\int\limits^{+\infty}_{-\infty}
d\nu \left(\frac{s}{s_0}\right)^{\bar \alpha_s(\mu_R) \chi(\nu)}
\alpha_s^2(\mu_R) c_1(\nu)c_2(\nu)\left\{1+\bar \alpha_s(\mu_R)
\left(\frac{c^{(1)}_1(\nu)}{c_1(\nu)}+\frac{c^{(1)}_2(\nu)}{c_2(\nu)}\right)
\right. \! \nonumber \\
&+&
\left.
\bar \alpha_s^2(\mu_R) \ln\left(\frac{s}{s_0}\right) \left[ \bar
\chi(\nu)+\frac{\beta_0}{8N_c}\chi(\nu)\left(-\chi(\nu)+\frac{10}{3}
+ i\frac{d\ln(\frac{c_1(\nu)}{c_2(\nu)})}{d\nu}+2\ln(\mu_R^2)  \right)
\right] \right. \nonumber \\
&+&
\Biggl.
\ln\left(\frac{s}{s_0}\right) \chi_{RG} (\nu) \Biggr\} \;. 
\end{eqnarray}
We find that
\beq{impsnu}
c_{1,2}(\nu)=
\frac{\left(Q^2_{1,2}\right)^{\pm i\nu-\frac{1}{2}}}{\sqrt{2}}
\frac{\Gamma^2 [\frac{3}{2}\pm i\nu]}{\Gamma [3\pm 2i\nu]}
\frac{6\pi}{\cosh (\pi
\nu)}\, ,
\eeq
\beq{c1c2s}
c_{1}(\nu)c_{2}(\nu)=\frac{1}{Q_1Q_2}\left(\frac{Q_1^2}{Q_2^2}\right)^{i\nu}
\frac{9\,\pi^3(1+4\nu^2)\sinh(\pi\nu)}{32\,\nu\,(1+\nu^2)\cosh^3(\pi\nu)} \, ,
\eeq
\beq{logratio}
i\frac{d\ln(\frac{c_1(\nu)}{c_2(\nu)})}{d\nu}=2\left[
\psi(3+2i\nu)+\psi(3-2i\nu)-\psi\left(\frac{3}{2}+i\nu\right)
-\psi\left(\frac{3}{2}-i\nu\right)-\ln\left(Q_1Q_2\right)
\right] \, .
\nonumber
\eeq

It can be useful to separate from the NLA correction to the impact
factor the terms containing the dependence on $s_0$ and $\beta_0$,
\bea{sepa}
C^{(1)}(\vec q^{\,\,2})&=&\int\limits^1_0 dz \,
\frac{\vec q^{\,\, 2}}{\vec q^{\,\, 2}+z \bar zQ^2}\phi_\parallel (z)
\\
&\times& \left[
\frac{1}{4}\ln\left(\frac{s_0}{Q^2}\right)\ln\left(\frac{(\alpha+z\bar
z)^4}{\alpha^2 z^2\bar z^2}\right)+\frac{\beta_0}{4N_c}\left(
\ln\left(\frac{\mu_R^2}{Q^2}\right)+\frac{5}{3}-\ln(\alpha)\right)
+\dots
\right] \;. \nonumber
\eea
Accordingly, one can write
\beq{ffss}
c^{(1)}_{1,2}(\nu)=\tilde c^{(1)}_{1,2}(\nu)+\bar c^{(1)}_{1,2}(\nu) \; ,
\eeq
where $\tilde c^{(1)}_{1,2}(\nu)$ are the contributions from the terms isolated
in the previous equation and $\bar c^{(1)}_{1,2}(\nu)$ represent the rest.
In Ref.~\cite{AlexDima1} it was found that
\bea{khk}
\frac{\tilde c^{(1)}_{1}(\nu)}{c_{1}(\nu)}+\frac{\tilde
c^{(1)}_{2}(\nu)}{c_{2}(\nu)}&=&\ln\left(\frac{s_0}{Q_1Q_2}\right)\chi(\nu)
+\frac{\beta_0}{2N_c}\left[\ln\left(\frac{\mu_R^2}{Q_1Q_2}\right)+\frac{5}{3}\right.
\nonumber \\
&+&
\left.
\psi(3+2i\nu)+\psi(3-2i\nu)-\psi\left(\frac{3}{2}+i\nu\right)
-\psi\left(\frac{3}{2}-i\nu\right)
\right] \, .
\eea

One can construct infinitely many representations of the amplitude, all of them equivalent within  
NLA accuracy. A particular one, motivated in Ref.~\cite{AlexDima2}, is to exponentiate 
all the scale-invariant part of the NLA kernel, obtaining 
\[
\frac{{\cal I}m_s\left( {\cal A} \right)}{D_1D_2}=\frac{s}{(2\pi)^2}
\int\limits^{+\infty}_{-\infty}
d\nu \left(\frac{s}{s_0}\right)^{\bar \alpha_s(\mu_R)
\chi(\nu)+\bar \alpha_s^2(\mu_R)
\left(
\bar
\chi(\nu)+\frac{\beta_0}{8N_c}\chi(\nu)\left[-\chi(\nu)+\frac{10}{3}
\right]
\right) + \chi_{RG} (\nu) }
\alpha_s^2(\mu_R) c_1(\nu)c_2(\nu)
\]
\beq{amplnlaE}
\times\! \left[1+\bar \alpha_s(\mu_R)
\left(\frac{c^{(1)}_1(\nu)}{c_1(\nu)}
+\frac{c^{(1)}_2(\nu)}{c_2(\nu)}\right)
+\bar \alpha_s^2(\mu_R)\ln\left(\frac{s}{s_0}\right)
\frac{\beta_0}{8N_c}\chi(\nu)\left(
i\frac{d\ln(\frac{c_1(\nu)}{c_2(\nu)})}{d\nu}+2\ln(\mu_R^2)
\right)\right].
\eeq
Another possible representation of the amplitude, in some sense closer to the 
original idea of the BFKL approach, is the ``series" representation, which reads
\bea{series}
\frac{Q_1Q_2}{D_1 D_2} \frac{{\cal I}m_s {\cal A}}{s} &=&
\frac{1}{(2\pi)^2}  \alpha_s(\mu_R)^2 \label{honest_NLA} \nonumber \\
& \times &
\biggl\{ b_0 + a_0\ln\left(\frac{s}{s_0}\right)
+\sum_{n=1}^{\infty}\bar \alpha_s(\mu_R)^n   \left[
a_n\ln\left(\frac{s}{s_0}\right)^{n+1} \right. \\
&+& \left.
 b_n \,
\biggl(\ln\left(\frac{s}{s_0}\right)^n   +
d_n(s_0,\mu_R)\ln\left(\frac{s}{s_0}\right)^{n-1}     \biggr) \right]
\biggr\}\;, \nonumber
\eea
where the coefficients
\beq{bs}
\frac{b_n}{Q_1Q_2}=\int\limits^{+\infty}_{-\infty}d\nu \,  c_1(\nu)c_2(\nu)
\frac{\chi^n(\nu)}{n!} \, ,
\eeq
are determined by the kernel and the impact factors in LLA and 
\beq{as}
\frac{a_n}{Q_1Q_2}=\int\limits^{+\infty}_{-\infty}d\nu \,  c_1(\nu)c_2(\nu) 
\chi_{RG} (\nu)
\frac{\chi^n(\nu)}{n!} 
\eeq
arise from the collinear improvement.
The coefficients
\[
d_n=n\ln\left(\frac{s_0}{Q_1Q_2}\right)+\frac{\beta_0}{4N_c}
\left(
(n+1)\frac{b_{n-1}}{b_n}\ln\left(\frac{\mu_R^2}{Q_1Q_2}\right)
-\frac{n(n-1)}{2} \right.
\]
\beq{ds}
\left.
+ \frac{Q_1Q_2}{b_n}\int\limits^{+\infty}_{-\infty}d\nu \, (n+1)f(\nu)
c_1(\nu)c_2(\nu)
\frac{\chi^{n-1}(\nu)}{(n-1)!}\right)
\eeq
\[
+\frac{Q_1Q_2}{b_n}\left(
\int\limits^{+\infty}_{-\infty}d\nu\, c_1(\nu)c_2(\nu)
\frac{\chi^{n-1}(\nu)}{(n-1)!}\left[
\frac{\bar c^{(1)}_{1}(\nu)}{c_{1}(\nu)}+\frac{\bar
c^{(1)}_{2}(\nu)}{c_{2}(\nu)}
 +(n-1)\frac{\bar \chi(\nu)}{\chi(\nu)}\right]
\right)
\]
are determined by the NLA corrections to the kernel and to the impact
factors. Here, $\bar c^{(1)}_{1,2}(\nu)$ represent the contribution without the 
terms depending on $s_0$ and $\beta_0$, and 
\beq{fv}
f(\nu)=\frac{5}{3}+\psi(3+2i\nu)+\psi(3-2i\nu)-\psi\left(\frac{3}{2}+i\nu\right)
-\psi\left(\frac{3}{2}-i\nu\right) \, .
\eeq
We stress that the terms in the series representation~(\ref{series}) with 
the $a_n$ coefficients are beyond the NLA, since, as one can easily see from 
Eq.~(\ref{bessel}), $\chi_{RG}$ is ${\cal O}(\bar \alpha_s^3)$.

\begin{figure}[tb]
\centering
\hspace{-1cm}
{\epsfysize 8cm \epsffile{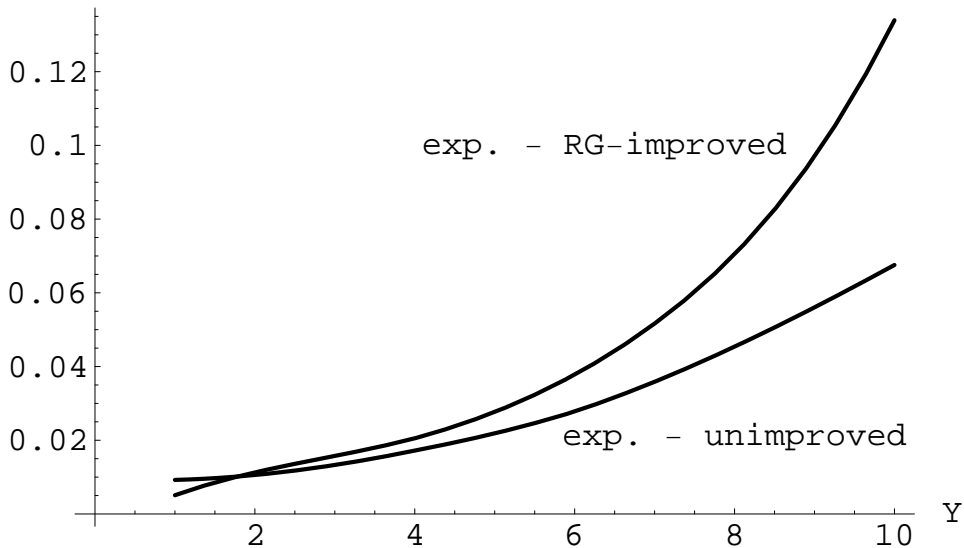}}
\caption[]{${\cal I}m_s ({\cal A})Q^2/(s \, D_1 D_2)$ as a function of
$Y$ at $Q^2$=24~GeV$^2$ and $n_f=5$ in the ``exponentiated" representation 
with and without collinear improvement of the kernel; in both cases the PMS 
optimization method has been used.}
\label{expRG}
\end{figure}

\section{Numerical results}

In this section we present some numerical results for the dependence in $s$ of the 
BFKL amplitude calculated for the process under study, using both the 
``exponentiated'' and the ``series'' representations derived in the previous
Section.
Following Ref.~\cite{AlexDima1}, we will adopt the principle of minimal 
sensitivity (PMS)~\cite{Stevenson} requiring, for each value of $s$, the minimal 
sensitivity of the predictions to the change of both the renormalization and the 
energy scale, $\mu_R$ and $s_0$. In previous studies, where the
unimproved kernel was used, the optimal choices for $\mu_R$ and $s_0$ turned out
to be very far from the kinematical scales of the process. Our aim is to see if 
and to what extent the inclusion of a collinear improvement leads to more ``natural''
values for the optimal scales. This would demonstrate that the RG-generated terms reproduce the essential subleading dynamics, thus stabilizing the perturbative series.
In the following analysis we use the two--loop running coupling
corresponding to the value $\alpha_s(M_Z)=0.12$.

\subsection{Symmetric kinematics}

We consider here the $Q_1=Q_2\equiv Q$ kinematics, i.e. the ``pure'' BFKL regime, 
with $Q^2$=24~GeV$^2$ and $n_f=5$. We start with the ``exponentiated" representation, given in Eq.~(\ref{amplnlaE}) and set $\ln(s/s_0)=Y-Y_0$, where $Y=\ln(s/Q^2)$ and $Y_0=\ln(s_0/Q^2)$.
We have looked for the optimal value for the scales $\mu_R$ and $Y_0$. In practice, for 
each fixed value of $Y$ we have determined the optimal choice of these parameters for 
which the amplitude is the least sensitive to their variation.
We have found that the amplitude is always quite stable under variation of both 
scales and exhibits generally only one stationary point (local maximum). We 
choose as optimal values of the parameters those corresponding
to this stationary point.

The optimal values turned out to be typically $\mu_R \simeq 3Q$ and 
$Y_0 \simeq 2$. In comparison with Ref.~\cite{AlexDima1}, where the optimal choice
was typically $\mu_R \simeq 10Q$, we can see that there is a remarkable move
towards ``naturalness''. The fact that the inclusion of the RG-terms affects 
the optimal choice of $\mu_R$ more strongly than of $Y_0$ is not surprising, since 
the added terms depend on $\mu_R$ and not on $Y_0$.
In Fig.~\ref{expRG} we show the result for the (imaginary part of the) 
``improved" amplitude compared with the result obtained in Ref.~\cite{AlexDima2}. 
The curves are in good agreement at the lower energies, the deviation increasing for 
large values of $Y$. This is consistent with having a larger asymptotic intercept when 
the collinear improvements are taken into account. We have to remember, however, that the applicability domain of 
the BFKL approach is determined by the condition $\bar \alpha_s(\mu_R) Y \sim 1$, 
that for our typical optimal value of $\mu_R$ and for $Q^2$=24~GeV$^2$ means 
$Y \sim 6$. Around this value the discrepancy is not so pronounced.

\begin{figure}[tb]
\centering
\hspace{-1cm}
{\epsfysize 8cm \epsffile{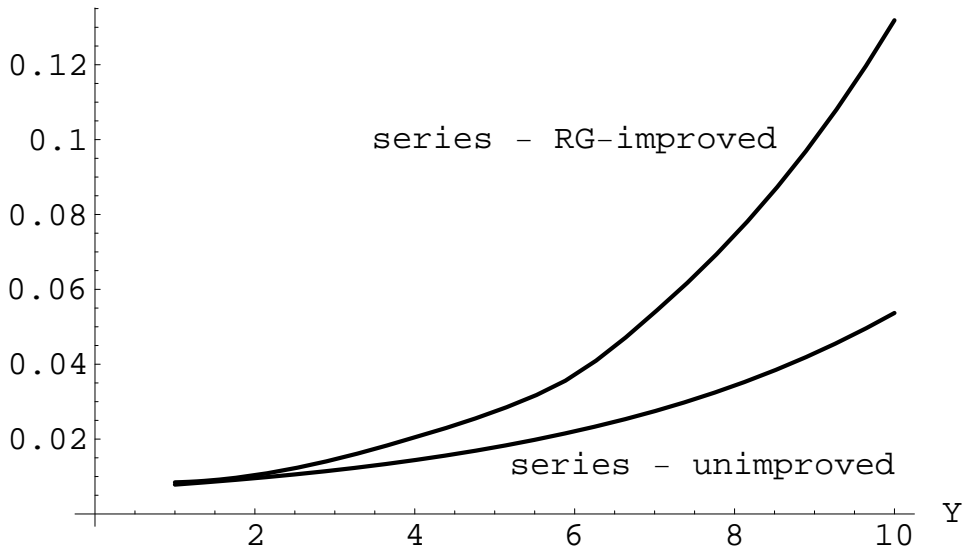}}
\caption[]{${\cal I}m_s ({\cal A})Q^2/(s \, D_1 D_2)$ as a function of
$Y$ at $Q^2$=24~GeV$^2$ and $n_f=5$ in the ``series" representation with and 
without collinear improvement of the kernel; in both cases the PMS 
optimization method has been used.}
\label{seriesRG}
\end{figure}

The next analysis has been done using the ``series" representation of the 
amplitude, given in Eq.~(\ref{series}). 
In this case we have also observed a smooth dependence of the amplitude on the 
two energy parameters. The optimal values for $Y_0$ and $\mu_R$ turned out to 
be quite similar to those obtained for the ``exponentiated"  representation, 
$\mu_R \simeq 3Q$ and $Y_0 \simeq 3$.
In Fig.~\ref{seriesRG} we show the behavior in $Y$ of the ``series" amplitude, compared with 
the determination of Ref.~\cite{AlexDima1}. The situation is similar to 
Fig.~\ref{expRG}, but the deviation between the curves appears to be more marked here.
It is important to observe that the curves for the ``exponentiated'' and  
``series'' representations of the amplitude as functions of $Y$ with collinear
improvement (see Figs.~\ref{expRG} and~\ref{seriesRG}) fall almost on top of each other, while in the 
determination without the collinear improvement there was a discrepancy, more 
pronounced at higher energies~\cite{AlexDima2}.
This is a further indication of a better stability, induced by the collinear
improvement.

In order to make visible the effect of the collinear improvement in the ``series''
representation we list the first few coefficients (see Eq.~(\ref{series})) $b_n$, 
$d_n$, coming from the unimproved BFKL kernel and impact factors 
(in LLA e NLA respectively), and $a_n$, coming from 
the RG-resummed terms. Using the optimal scales chosen with the PMS method we obtain
($Q^2$=24 GeV$^2$, $n_f=5$, $Y_0=3$, $\mu_R=3 Q$)
\beq{coe}
\begin{array}{lllll}
b_0=17.0664  & b_1=34.5920   & b_2=40.7609  & b_3=33.0618   &
b_4=20.7467  \\
& & & & \\
& d_1=0.674275 & d_2=-1.73171 & d_3=-7.46518 & d_4=-15.927 \\ 
& & & & \\
& a_1=5.52728 & a_2=7.30295 & a_3=6.42149 & a_4=4.24011 \;. \\
\end{array}
\eeq

We can see that the $a_n$ coefficients are of the opposite sigh respect to the 
$d_n$, so ``curing" the bad behavior of the BFKL series. Even if the values of 
the $a_n$ coefficients go down with $n$, they appear in Eq.~(\ref{series}) 
with two more powers of the energy logarithm than the $d_n$ coefficients, so that
their effect is not limited to low energies.

\begin{figure}[tb]
\centering
\hspace{-1cm}
{\epsfysize 8cm \epsffile{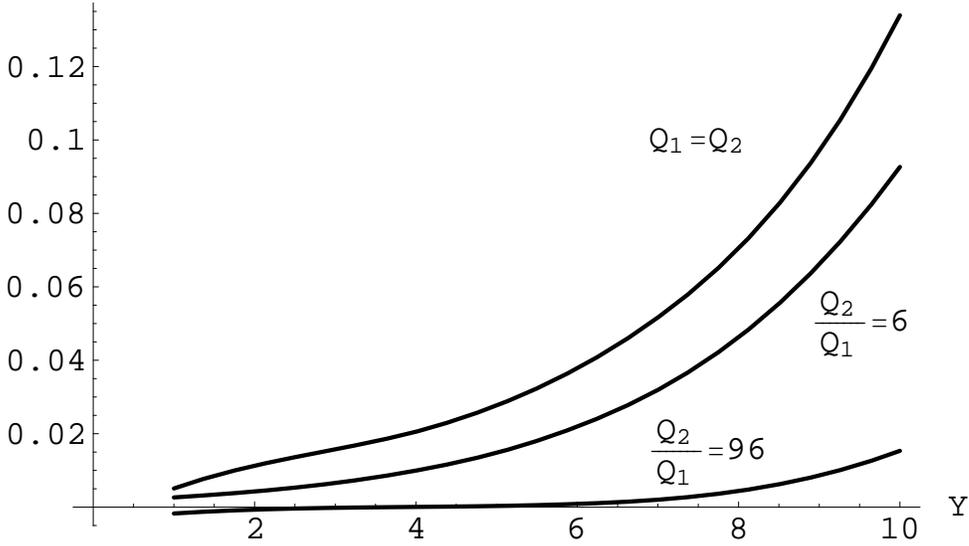}}
\caption[]{${\cal I}m_s ({\cal A})Q_1 Q_2/(s \, D_1 D_2)$ as a function of $Y$ for photons
with strongly ordered virtualities ($Q_2/Q_1=6$ and $Q_2/Q_1=96$, with $Q_1 Q_2$=24~GeV$^2$), 
in comparison with the case of photons with equal virtualities ($Q_1^2=Q_2^2$=24~GeV$^2$).
All curves have been obtained using the ``exponentiated" representation with the 
collinearly improved kernel.}
\label{Asymmtot}
\end{figure}

\subsection{Asymmetric kinematics}

When the virtualities of the photons are strongly ordered, we enter 
the ``DGLAP'' regime, where collinear effects should come heavily into the game.
In this regime, previous attempts to numerically determine the amplitude using
unimproved kernels were unsuccessful due to severe instabilities~\cite{AlexDima3}.
We have found here that these instabilities disappear if, instead, the 
RG-improved kernel is used.

In the numerical analysis to follow, we consider two choices for the 
virtualities of the photons, $Q_1$=2~GeV, $Q_2$=12~GeV and $Q_1$=0.5~GeV,
$Q_2$=48~GeV, so that $Q_1 Q_2 = Q^2$=24 GeV$^2$ in both cases, and
used the ``exponentiated" representation. We define $Y=\ln(s/Q_1 Q_2)$  
and $Y_0=\ln(s_0/Q_1 Q_2)$. 

For the first choice of virtualities, we find that for each $Y$ value 
the amplitude is still quite stable under variation of the energy parameters and 
the optimal values are $\mu_R \simeq 4\sqrt{Q_1 Q_2}$ and $Y_0 \simeq 2$,
almost independently of $Y$. The same holds for the second choice of virtualities, 
with the only difference that now the optimal values depend strongly on $Y$.
As an example, for $Y=6$, when $\bar \alpha_s(\mu_R) Y \sim 1$, the optimal 
$\mu_R$ is $\simeq 3\sqrt{Q_1 Q_2}$, but $Y_0$=7. This large value for $Y_0$
should not be surprising: if we use $Q_2^2$ as normalization scale in $Y_0$
instead of $Q_1 Q_2$, the optimal value lowers down $\sim $2.5, which looks
more ``natural''.

In Fig.~\ref{Asymmtot} we plot the amplitude for the two choices of 
photons' virtualities we have considered, together with the amplitude for
$Q_1=Q_2=\sqrt{24}$ GeV. The amplitude becomes smaller and smaller when 
$Q_2/Q_1$ increases, as it must be expected due to the presence of the 
factor $\cos(\nu \log(Q_2^2/Q_1^2))$ in the integration over $\nu$. We stress
again that, if the RG-generated terms are removed, it is impossible even to draw
the curves in Fig.~\ref{Asymmtot} with $Q_2\neq Q_1$.

\section{Conclusions}

We have applied a RG-improved kernel to determine the amplitude for the forward 
transition from two virtual photons to two light vector mesons in the Regge 
limit of QCD with next-to-leading order accuracy. 
The result obtained is independent on the energy scale $s_0$, and on the
renormalization scale $\mu_R$ within the next-to-leading approximation.

Using two different representations of the amplitude, which include the dependence
on the energy scale and on the renormalization scale at subleading level,
we have performed a numerical analysis both in the kinematics of
equal and strongly ordered photons' virtualities.

An optimization procedure, based on the principle of minimal sensitivity,
has led to results stable in the considered energy interval, which allow to 
predict the energy behavior of the forward amplitude. The important finding is that 
the optimal choices of $s_0$ and $\mu_R$ are much closer to the kinematical
scales of the problem than in previous determinations based on unimproved kernels. 
This effect is very marked for $\mu_R$, as it must be expected, since the extra-terms
depend on $\mu_R$ and not on $s_0$. 
This leads us to conclude that the extra-terms in the BFKL kernel coming from 
collinear improvement, which are subleading to the NLA, catch an important 
fraction of the dynamics at higher orders.

Moreover, the use of the improved kernel has allowed to obtain the 
energy behavior of the forward amplitude in the case of strongly ordered  
photons' virtualities, which turned out to be unaccessible to previous attempts
using unimproved kernels.

\vspace{1.0cm} \noindent
{\Large \bf Acknowledgment} \vspace{0.5cm}

We thank D.Yu.~Ivanov for reading the manuscript and making valuable comments.

\end{document}